\documentstyle[aasms]{article}

\def\lae{\mathrel{<\kern-1.0em\lower0.9ex\hbox{$\sim$}}}
\def\gae{\mathrel{>\kern-1.0em\lower0.9ex\hbox{$\sim$}}}

\font\fsmall=cmr10

\def\farcm{\hbox{$.\mkern-4mu^\prime$}}
\def\farcs{\hbox{$.\!\!^{\prime\prime}$}}

\journalid{337}{15 January 1989}
\articleid{11}{14}

\slugcomment{To appear in the Astronomical Journal}

\begin{document}

\title{Spectroscopic Binaries in Globular Clusters. I. A Search for Ultra-Hard Binaries on the 
Main Sequence in M4\altaffilmark{1}}

\author{Patrick C\^ot\'e\altaffilmark{2}}
\affil{Department of Physics and Astronomy, McMaster University, Hamilton, ON, L8S 4M1, Canada\\and\\
    Dominion Astrophysical Observatory, Herzberg Institute of Astrophysics,
    National Research Council of Canada, 5071 W. Saanich Road, R.R.5, Victoria, BC, V8X 4M6, Canada\\
    cote@dao.nrc.ca}

\and

\author{Philippe Fischer\altaffilmark{2,3}}
\affil{AT\&T Bell Laboratories, 600 Mountain Avenue, 1D-316, Murray Hill, NJ 07974
\\and\\
Department of Astronomy, University of Michigan, Dennison Building, Ann Arbor, 
MI 48109-1090\\philf@astro.lsa.umich.edu}


\altaffiltext{1}{Based in part on observations collected at Kitt Peak National Observatory,
National Optical Astronomy Observatories, operated by the Association of
Universities for Research in Astronomy, Inc., under contract with the
National Science Foundation.}
\altaffiltext{2}{Visiting Astronomer at the Cerro Tololo Inter-American Observatory,
National Optical Astronomy Observatories, operated by the Association of
Universities for Research in Astronomy, Inc., under contract with the
National Science Foundation.}
\altaffiltext{3}{Hubble Fellow.}


\begin{abstract}
A search for spectroscopic binaries on the main sequence of the nearby globular cluster
M4 has been undertaken with Argus, the multi-object spectrograph on the CTIO 4.0m telescope.
A pair of radial velocities (median precision $\simeq$ 2 km s$^{-1}$) separated by 
11 months have been obtained for 33 turnoff dwarfs in the magnitude range 16.9 $\le$ V $\le$ 17.4. 
Monte-Carlo simulations have been used to derive a binary fraction, $x_b$,
for systems with periods in the range 2 days $\lae {\rm P} \lae$ 3 years and 
mass ratios between 0.2 and 1.0. This short-period cutoff is more than an order of magnitude 
smaller than those of existing radial velocity surveys and is comparable to the shortest periods 
possible for main-sequence turnoff stars. Our survey therefore provides a first glimpse into the 
abundance of ``ultra-hard" spectroscopic binaries in globular clusters. 
Although no star shows a velocity variation larger than 14 km s$^{-1}$, two objects
are observed to have chi-square probabilities below 0.1\%. No such stars are expected
in a sample of 33.
We find a best-fit binary fraction of $x_b \simeq$ 0.15,
a value which is consistent with recent estimates based on deep
HST color-magnitude diagrams, as well as with the binary fraction of $x_b$ $\simeq$ 
0.1 for nearby, solar-type stars having similar periods and mass ratios.
Our derived binary fraction suggests that exchange interactions with pre-existing binaries
are a plausible means of explaining the origin of the hierarchical triple system containing the pulsar PSR 1620-26.

\end{abstract}


%
%

\section{Introduction}
It is now clear that the dynamical evolution of globular clusters is intimately 
tied to their binary content. Perhaps the most dramatic consequence of a population
of primordial binaries is their tendency to delay the onset of core collapse (Spitzer \& Mathieu 1980).
Like most other Galactic globular clusters, M4 shows no evidence for a
central brightness cusp (Trager, King \& Djorgovski 1995), despite the fact that the cluster age 
(Richer \& Fahlman 1984) exceeds its central relaxation time (Djorgovski 1993) by a factor of approximately four
hundred.
On this basis, M4 should now be in the post-collapse phase of its evolution (Lightman, Press \& Odenwald 1978).
The lack of an observable central surface density cusp may therefore be indirect evidence that
the cluster contains some binaries.
More compelling evidence for an appreciable binary fraction, $x_b$, comes from the existence of the 
hierarchial triple pulsar system
PSR 1620-26, the first triple star ever detected in a globular cluster. Formation models for this
exotic system (e.g., Rasio, McMillan \& Hut 1995) generally 
invoke exchange interactions between pre-existing binaries and, consequently, require a significant binary
fraction. 

However, measuring the globular cluster binary fraction has proven to be a difficult task.
Efforts to do so have generally proceeded along three lines: (1) searches
for radial velocity variables (Pryor, Latham \& Hazen 1988; Hut et al. 1992; C\^ot\'e et al. 1994); 
(2) photometric searches for eclipsing binaries (Mateo et al. 1990; Yan \& Mateo 1994) or
cataclysmic variables (Shara et al. 1995) and;
(3) measurements of the so-called ``second sequence" in the cluster color-magnitude diagram 
(Romani \& Weinberg 1991). 
Although this last technique has the important advantage that it is a single-epoch 
observation, it is complicated by the effects of reddening and/or metallicity variations,
and requires a profound understanding of the photometric uncertainties. 
Moreover, it is fundamentally incapable of probing the distribution of orbital
elements since it is impossible to distinguish between systems with vastly different
orbital periods and requires an {\sl a priori} assumption about the distribution of mass ratios.
While searches for eclipsing binaries {\sl are} able to measure the distribution of orbital
elements, they are almost completely insensitive to systems with periods longer than
$\sim$ 10 days (e.g., Mateo 1993).
On the other hand, radial velocity surveys are capable of identifying
binaries over a much wider range in orbital period and, through continued monitoring, of 
determining directly the distributions of orbital elements. 

Until recently, limited instrumental efficiency confined searches for radial velocity variables in globular 
clusters to the brightest red giants. Monte-Carlo simulations show that 
the binary discovery efficiency near the tip of the giant branch is only a few percent, so that 
many stars must be surveyed in order to place meaningful constraints on the binary fraction.
To further complicate matters, stars near the tip of the giant branch occasionally exhibit
atmospheric pulsations (Gunn \& Griffin 1979; Mayor et al. 1984) which
mimic the effects of unseen companions.
In addition, it is now recognized that selection effects caused by Roche-lobe overflow
in these evolved systems are potentially severe.
Since the likelihood of mass transfer between the binary components is strongly dependent 
on their separation at periastron, it is the ``hardest" systems 
which are most likely to be missed in magnitude-limited samples
(i.e., those systems which are expected to play the most important role in the dynamical 
evolution of the cluster).
This effect, combined with the limited time coverage of existing observations,
means that most radial velocity surveys of globular clusters are poorly suited to identifying
binaries with periods less than $\sim$ 0.2 year or greater than $\sim$ 20 years.
The two-fold strategy required to discover binaries outside this rather narrow range of period is clear:
(1) survey a large number of main-sequence stars in order to identify the short-period 
systems; and (2) continue to monitor the radial velocities of red giants in order to detect long-period binaries. 
In this paper, we present the first results of a program to measure $x_b$ based 
on radial velocities of main-sequence stars in the globular cluster M4. In a companion
paper, we report on a search for long-period binaries among the red giants of M22 using
a sample of radial velocities which spans more than two decades (C\^ot\'e et al. 1996).

M4 is the nearest globular cluster and, at a distance of only $\simeq$ 2.0 kpc (Djorgovski 1993), it is
the obvious target for a search for ``ultra-hard", main-sequence 
binaries.\altaffilmark{4}\altaffiltext{4}{Although Djorgovski (1993) quotes a distance of 1.5 kpc for Terzan 6, 
Fahlman, Douglas \& Thompson (1995) have recently used JHK$^\prime$ photometry to derive an improved estimate 
of 6.8 kpc.}
Although it is heavily reddened, the main-sequence turnoff appears at V $\simeq$ 16.8 which makes
high-resolution, spectroscopic observations of turnoff dwarfs feasible with 4m-class telescopes.
In addition, it has a relatively high metallicity of [Fe/H] = --1.1 (Drake, Smith \& Suntzeff 1992), an important
feature for measuring precise radial velocities at low S/N levels.
Using Argus, the multi-object spectrograph on the CTIO 4m telescope, we have measured a pair of
radial velocities separated by 11 months for 33 stars near the cluster main-sequence turnoff.
Monte-Carlo simulations demonstrate that these observations are
capable of detecting binaries with periods between 2 days and 3 years, a span of 2.75 decades in period.
This represents a $\sim$ 40\% increase in the sampling of the binary period distribution function
over surveys with more extensive time coverage (e.g., Hut et al. 1992; C\^ot\'e et al. 1994).
Our short-period detection limit of $\simeq$ 2 days (see \S 3) is also comparable to the shortest periods possible for
main-sequence turnoff stars and therefore provides a first glimpse into the abundance of 
ultra-hard spectroscopic binaries in any globular cluster.

\section{Observations}
\subsection{Photometry}

A pair of wide-field BV images of M4 (kindly donated by Tad Pryor) were used to construct 
a color-magnitude diagram (CMD) from which candidate cluster dwarfs were selected.
Both images were obtained during new moon on June 2 1992 with the KPNO 
0.9m telescope and 2048$\times$2048 T2KA CCD 
(scale = 0$\farcs$689 pixel$^{-1}$, gain = 4.2 e$^-$ ADU$^{-1}$ and readnoise = 4 e$^-$).
Since the cluster was centered on both frames, stars were measured out to a maximum
distance of approximately 16$'$ from the Shawl \& White (1986) cluster center 
(see Figure 1 for a reproduction of the V image with our 
program objects identified according to the numbering scheme of Table 1). 
Exposure times were 160s for V and 300s for B. In both cases, the FWHM of isolated stars was 
$\simeq$ 2$\farcs$0.
After standard preprocessing (i.e., bias subtraction, trimming, flat-fielding) with 
IRAF,\altaffilmark{5}\altaffiltext{5}{IRAF is distributed by the National Optical
Astronomy Observatories, which are operated by the Association of Universities for
Research in Astronomy, Inc., under contract to the National Science Foundation.}
an instrumental CMD was derived using DAOPHOT II (Stetson, Davis \& Crabtree 1990) and
calibrated using uncrowded stars chosen from Alcaino, Liller \& Alvarado (1988).
The resulting CMD is shown in Figure 2 where the dotted lines 
at V = 16.9 and 17.4 indicate the magnitude interval from which stars were randomly selected
for subsequent spectroscopy.
Since accurate astrometric positions
are required by Argus, we used 41 stars from the HST Guide Star Catalog to
derive positions good to $\simeq$ 0$\farcs$5 for our program stars.

\subsection{Spectroscopy}

High-resolution spectra were accumulated using Argus, the multi-object spectrograph
on the CTIO 4.0m telescope. This instrument consists of 24 independently controlled fiber positioning arms
mounted around the 45$^\prime$-diameter prime focus field. 
The 1$\farcs$9-diameter fibers carry
light from the prime focus cage to a spectrograph located in a thermally and mechanically
isolated room in the 4.0m dome. We used the 31.6 lines mm$^{-1}$ echelle grating and
CTIO \#11 order-selection filter to isolate the spectral region 5090 -- 5260~\AA. This region contains
the Mg I triplet as well as numerous other sharp absorption lines and is known to be ideally suited to
the measurement of accurate radial velocities at low S/N levels (see C\^ot\'e et al. 1991).

All of the radial velocities presented here were obtained during two observing
runs with Argus on July 14--16 1993 and June 8--10 1994.
For both observing runs, a Reticon 400$\times$1200 CCD (binned by a factor of two in the spatial
direction) was used to record the spectra.
Object exposure times were 3600s.
Th-Ar comparison lamp spectra were obtained every two or three hours throughout the night.
High-S/N spectra for both the twilight/dawn sky and
several IAU radial velocity standard stars were also collected, although subsequent experimentation
demonstrated that the co-added sky spectrum produced the best cross-correlations and was 
therefore adopted as the final template. 
After the images were trimmed and corrected for overscan and bias,
object spectra were identified, traced and extracted using the APSUM task
in IRAF. Once corrected for cosmic rays and continuum-subtracted, 
the extracted spectra were divided by an identically extracted
(and normalized) quartz lamp flat-field to remove pixel-to-pixel variations. Individual lines in a 
single Th-Ar lamp spectrum were then identified and used as a reference to re-identify the
remaining lamp spectra. Typically, 14 -- 16 features were identified in each aperture
with an RMS error in the dispersion-solution of $\simeq$ 0.01 \AA .
The object spectra were then dispersion-corrected
in the range 5090 -- 5260 \AA\ and cross-correlated using the
FXCOR task in IRAF. A ramp filter was used in all cases to filter 
both the object and template spectra. 
The sample of stars having repeat velocities is limited by our July 1993 run: due to poor
weather conditions, reliable cross-correlations were obtained for only 35 stars.
Conditions were better during our June 1994 observing run where we measured velocities
for a total of 173 stars. In what follows, we restrict our discussion to those
35 stars which were observed during both runs (i.e., 33 cluster dwarfs, one cluster red giant and one nonmember).

Naturally, misidentifications at the telescope will compromise any attempt to measure the binary fraction.
Since Argus is known to have difficulty in positioning fibers at the $\lae$ 1$^{\prime\prime}$ level, such 
misidentifications are a potentially serious concern for main-sequence turnoff stars in crowded globular cluster fields. 
Prior to each field setup, we therefore used the Argus periscope to check the selection of our program
objects against periscope finder charts generated in advance with the KPNO images.
We are therefore certain that all objects have been properly identified.
A second, related, concern is potential contamination of the target spectra by close, {\sl optical} 
companions (i.e., small differences in fiber centering between runs might give rise to spurious 
variability). Since the likelihood of contamination depends, of course, on the local surface density, 
we have limited our target selection to those dwarfs which are more than 3$\farcm$2 from the cluster
center (Shawl \& White 1986). This corresponds to $\simeq$ 0.7$r_h$ (Trager, King \& Djorgovski 1995),
so that the bulk of our program objects lie at, or beyond, the cluster half-light radius.
Our survey therefore provides no information on binaries which may reside in the cluster core.
On the basis of mass segregation alone, one would expect a preponderance of binaries in
the central regions of the cluster. However, the steep gradient in the distribution of binaries
expected from energy equipartition may be partly erased by destructive encounters with other 
stars (i.e., the binary disruption timescale at the cluster center is much shorter than at the half-light radius).
Moreover, interactions in the cluster core are expected to eject some binaries to the cluster envelope
and may lead to an overabundance of hard binaries at large radii (Sigurdsson \& Phinney 1995).

Any search for radial velocity variables requires a firm understanding of the observational errors. 
Since experience has shown that FXCOR tends to overestimate the true
uncertainties, we intended to use repeat measurements from each
run to bootstrap our FXCOR estimates to the true error distribution. Unfortunately,
a conspiracy of problems with the weather, CCD controller and unexpectedly large variations in
the fiber-to-fiber throughput limited the number of such
measurements. In total, we have repeat velocities for only six stars in M4,
although, as Figure 3 demonstrates, our adopted uncertainties 
appear to be reasonable (and the empirical scaling factor derived from these stars is in good
agreement with that found for red giants in the several other globular clusters which were
observed during these same runs).
Another way to test whether our adopted uncertainties are realistic is to examine the distribution of
P($\chi^2$), the chi-square probabilities (see the upper panel of Figure 4). 
For a sample of constant velocity stars, the histogram 
should appear flat, whereas the distribution for a sample which includes a number of radial velocity variables 
will show an excess near P($\chi^2$) $\approx$ 0.
Although our sample is rather small, we see that the P($\chi^2$) histogram appears 
suitably flat with no obvious skew toward low probabilities.
The lower panel of Figure 4 shows the same data plotted as a cumulative distribution (this 
technique avoids the problem --- inherent in the above approach --- of choosing an optimum bin width).
The distribution expected from a sample of constant velocity stars having correctly estimated uncertainties
is given by the diagonal line. Once again, the strong correspondence between the two distributions
lends credence to our adopted uncertainties.

The complete sample of 34 member stars has a combined $\chi^2$ of 67.0 for 39 degrees of freedom.
Although this is much larger than expected, the $\chi^2$ drops to 33.6 for 37 degrees of freedom
if we exclude the two stars which show velocity differences of more than 10 km s$^{-1}$ and have P($\chi^2$) $<$ 0.001. 
This corresponds to P($\chi^2$) = 0.63, which we consider acceptable, and conclude that our adopted 
uncertainties are representative of the true error distribution.
The final data are presented in Table 1 which records the
object identification number, distance from the cluster center and position angle, the
heliocentric Julian date of the observation, the measured velocity, 
the number of velocities, the weighted mean velocity, the chi-squared, the probability P$(\chi^2)$ of that 
chi-squared value being exceeded assuming that the velocity is constant, the V magnitude, and B-V color.
Based on the data presented in Table 1, we find a systemic
radial velocity for M4 of 70.3$\pm$0.7 km s$^{-1}$ using the maximum-likelihood estimator of
Pryor \& Meylan (1993). This is consisent with the value of 70.9$\pm$0.6 km s$^{-1}$ reported
by Peterson et al. (1995) from accurate velocities for 180 cluster giants.

\section{Monte-Carlo Simulations} 

Only certain {\sl types} of binaries (i.e., those in a limited range of orbital period P and mass ratio, 
q = M$_2$/M$_1$) will be discovered in any radial velocity survey.
As a result, estimates of $x_b$ should refer explicitly to these ranges.
In this section, we describe the models used to extract $x_b$ from our radial velocities and consider 
in detail the range in orbital periods and mass ratios to which our observations are sensitive.
Such {\sl binary discovery efficiencies} depend on a variety of parameters including the number, spacing and precision
of the radial velocities, as well as the details of the assumed binary population. 
Our procedure for generating a simulated set of radial velocities is described below
(see also Hut et al. 1992 and C\^ot\'e et al. 1994). By comparing the actual catalog of radial velocities to 
numerous simulated ones, each of which contains a known number of binaries, we are able to determine both the 
best-fit binary fraction and its corresponding confidence limits.

We begin by choosing a binary fraction and, based on this value, randomly assigning either 
binary or single star status to each star in the survey.
We then compute the radius, R, of each primary by combining
the [Fe/H] = --1.26, [O/Fe] = +0.55, 15 Gyr isochrone of Bergbusch \& VandenBerg (1992) with the photometry
given in Table 1. For each object, we interpolated in radius, although the difference in R between our
brightest (V = 16.90) and faintest (V = 17.37) dwarfs is nearly negligible (i.e., R decreases from 1.19 to 0.96R$_{\odot}$,
assuming V(HB) = 13.35 and $M_{\rm V}$(HB) = +0.6).
Once an estimate of the radius is in hand, we assign orbital elements to the binary. This involves selecting the 
longitude and time of periastron passage, inclination, eccentricity, orbital period and component masses.
While the first three parameters may be drawn at random, some care must be taken in choosing the last
three parameters since virtually nothing is known about their distribution in globular clusters.

In their comprehensive survey of multiplicity among nearby, solar-type stars, Duquennoy \& Mayor (1991)
found that the distribution of binary orbital periods can be well approximated by 
$${d{\rm N} \over d{\rm \log{P}}} \propto \exp(-{(\log{\rm P} - \overline{\log{\rm P}})^2 \over 
{2\sigma_{\rm log~P}^2}}),\eqno{(1)}$$
where P is in days, $\overline{\log{\rm P}}$ = 4.8 and $\sigma_{\rm \log{P}}$ = 2.3. We have used
this distribution to select orbital periods in our simulations.
Likewise, for the distribution of mass ratios, we adopt the Duquennoy \& Mayor (1991) distribution
$${d{\rm N} \over d {\rm q}} \propto \exp(-{({\rm q} - \overline{\rm q})^2 \over {2\sigma_{\rm q}^2}}),\eqno{(2)}$$ where 
$\overline{\rm q} =$ 0.23 and $\sigma_{\rm q}$ = 0.42.
Based on the Bergbusch \& VandenBerg (1992) models, we take the mass
of the primary to be ${\rm M_1} = 0.81$M$_{\odot}$. Mass ratios were randomly chosen from the interval
0.2 $\le$ q $\le$ 1.0, so that the M$_2$ is confined to the range 0.16 -- 0.81M$_{\odot}$. 
In \S 4, we discuss the effects of different input parameters on the derived binary fraction 
(for instance, including secondaries
with masses near the hydrogen-burning limit or adopting distributions which are flat in log P and log q).

Assumptions concerning the distributions of eccentricities have potentially more serious consequences for
the derived binary fractions (see Figure 1 of Hut et al. 1992). We have carried out simulations for three different
assumed distributions: (1) purely circular orbits, $e=0$; (2) a thermal distribution of eccentricities, $f(e) = 2e$ 
(expected if the distribution is a function of energy only, a condition thought to have been established
through tidal encounters; Heggie 1975) and; (3) a distribution of eccentricities like that seen
in samples of halo binaries (Latham et al. 1992), 
namely an $f(e) = 2e$ distribution for periods above $\simeq$ 19 days
and circular orbits below.\altaffilmark{6}\altaffiltext{6}{Such a distribution is thought to be the result of
tidal circularization forces acting on the binary components for 
$\sim$ 15 Gyr.} For red giants like those studied in previous surveys of this
sort (e.g., Hut et al. 1992), the prevalence of mass transfer among high eccentricity systems results in a 
markedly lower discovery efficiency compared to circular systems.
The reduced likelihood of mass transfer among our sample of turnoff dwarfs (as well as our sensitivity to
just the shortest period binaries) means that, to first order,
we are equally likely to detect binaries drawn from any of our three different eccentricity distributions
(see below).

We test for mass transfer by checking to ensure that each binary has a separation at closest 
approach  which exceeds 
$$a_{\rm crit} = {\rm R}/h({\rm q}),\eqno{(3)}$$
where $h({\rm q})$ is a slowly varying function which goes from 0.38 to 0.59 as q changes
from 1 to 1/12 (Pryor, Latham \& Hazen 1988). If the periastron distance is less than $a_{\rm crit}$, 
we assume that the binary will have undergone a merger and be undetectable (i.e., its orbital angular momentum
will have been transformed into axial rotation of the primary) and the selection process is repeated.
Each simulated velocity includes a realistic amount of observational noise, which we estimate from the
actual velocity uncetainties.
Compared to previous surveys, which monitored just the brightest and most highly evolved cluster stars,
the importance of mass transfer is greatly reduced for our sample of turnoff dwarfs which have
{\sl much} smaller radii (e.g., R $\approx$ 1R$_{\odot}$ as opposed to 11R$_{\odot}$ $\lae$ R $\lae$ 82R$_{\odot}$ for 
the M22 red giants monitored by C\^ot\'e et al. 1996).
For each $x_b$, we generate 1000 simulated data sets consisting of 73 radial velocities 
for 34 members stars. This procedure is
carried out for a grid of $x_b$ ranging from
0.0 to 0.6 in increments of ${\Delta}x_b = 0.01$.

Before comparing the simulations to the actual velocities, we must determine the fraction of
{\sl known} binaries which we expect to discover with these observations.
Since a given set of radial velocities is only capable of revealing binaries within certain ranges of orbital 
period and mass ratio, some care must be exercised to ensure that appropriate upper and lower limits 
for the distribution of these parameters are used in generating the simulations. 
In Figure 5 we present binary discovery efficiencies as a function of orbital period
for systems in the range 1 day $\le$ P $\le$ 10 years and 0.2 $\le$ q $\le$ 1.0 based on
the data given in Table 1. 
We consider a binary ``discovered" if it shows a velocity difference greater than 10 km s$^{-1}$
(corresponding to P($\chi^2$) = 0.0013 for our median velocity uncertainty of 2.2 km s$^{-1}$).
The discovery efficiencies for the cases of
circular orbits, thermal orbits and halo orbits are shown as the solid, dashed and dotted lines, respectively.
For each eccentricity distribution we present two curves: the upper one shows the binary 
discovery efficiency {\sl before} taking into account Roche-lobe overflow, while the lower
curve shows the same quantity {\sl after} the effects of mass transfer are included.
Although our time coverage is much more limited than published surveys (e.g., 11 months compared
to 22 years for the survey of M22 red giants by C\^ot\'e et al. 1996), we conclude from Figure 5 that
the present survey should be capable of detecting binaries with 2 day $\lae$ P $\lae$ 3 years
and 0.2 $\lae$ q $\lae$ 1.0. This range of 2.75 decades in $\log$ P is larger than that for
any published radial velocity survey of globular cluster stars.

Two different comparisons are used to find the best match between the actual and simulated 
radial velocities. First, for each $x_b$ we calculate the modal number of stars which exhibit velocity differences 
greater than 8 and 10 km s$^{-1}$. These thresholds were chosen since they represent approximate 4- and 5-$\sigma$ 
deviations with a median velocity uncertainty of $\sim$ 2 km s$^{-1}$. It is worth emphasizing that 
the atmospheric motions of up to $\sim$ 8 km s$^{-1}$ which have plagued radial velocity surveys of red giants
(Gunn \& Griffin 1979; Mayor et al. 1984; C\^ot\'e et al. 1996)
are avoided by studying main-sequence stars.
The median value of $x_b$ which produces an identical number of large velocity differences to the
actual data is then adopted as the best-fit binary fraction. Upper and lower confidence limits 
on $x_b$ are then estimated directly from the simulations.
Second, we compare the cumulative distribution of observed velocity differences for the entire sample of 34 members
to the mean distribution of simulated differences for each binary fraction.
For those stars which have more than two measurements, 
we use only the largest velocity difference in order to ensure that the differences
are independent and that a comparison to the true cumulative distribution via a Kolmogorov-Smirnov
test is appropriate (Pryor, Latham \& Hazen 1988).

\section{Results}

Two of our 33 dwarfs (stars \#9 and \#23) show velocity differences greater than 10 km s$^{-1}$ and
have P($\chi^2$) $\le$ 0.1\%. (Note that a third star --- object \#5 --- may also shows some evidence for binarity: a 
velocity variation of 9.15 km s$^{-1}$ and  P($\chi^2$) = 0.3\%.) Inspection of Figure 5
reveals that, for these data, our binary discovery efficiency is $\sim$ 0.35 for binaries with periods
in the range 2 days $\lae {\rm P} \lae$ 3 years and mass ratios between 0.2 and 1.0. 
A crude estimate of the actual binary fraction based on this detection fraction of 2/33 = 0.06
is thus $x_b$ $\sim$ 0.17. 
A more exact estimate of $x_b$ is given below.

Based on the observed number of velocity differences greater than 8 and 10 km s$^{-1}$, our best estimates
for the cluster binary fraction are $x_b$ = 0.18$^{+0.29}_{-0.18}$ (circular orbits), 
0.23$^{+0.34}_{-0.23}$ (thermal orbits) 
and 0.19$^{+0.34}_{-0.19}$ (halo orbits), where the uncertainties refer to 90\% confidence limits.
Although these ranges are clearly larger than is desirable, we can
also estimate $x_b$ by comparing the cumulative distribution of velocity differences 
for {\sl all} stars in the survey.
This comparison has the advantage that it makes use of the entire sample of stars with
multiple measurements (i.e., not just those stars which show large velocity variations) although it does
require that the distribution of measurement errors be well understood.

Figure 6 shows the cumulative distribution of velocity differences for the three cases of 
circular, thermal and halo orbits. The solid
lines depict the distribution of measured velocity differences; the dotted lines indicate,
from top to bottom, the distributions for simulations containing 0, 8, 16, 24, 32 and 40\% binaries. 
Based on a Kolmogorov-Smirnov test, we conclude that the 
best match between the data and simulations is found for $x_b$ = 0.07 (circular orbits), 
0.09 (thermal orbits) and 0.08 (halo orbits).
The upper limits determined using this approach are more stringent than those
obtained by simply counting the number of large velocity differences. For the three cases of circular,
thermal and halo orbits, we find respective 90\% confidence upper limits on $x_b$ of
0.26, 0.32 and 0.30. Taking a straight average of the binary fractions derived using the two different
approaches gives $x_b$ of 0.13$^{+0.13}_{-0.13}$ (circular orbits), 0.16$^{+0.16}_{-0.16}$
(thermal orbits) and 0.14$^{+0.16}_{-0.14}$ (halo orbits), which we adopt as our best estimates of the M4 
binary fraction for systems with 2 days $\le {\rm P} \le$ 3 years and 0.2 $\le {\rm q} \le$ 1.0.
Although the allowed range of binary fractions based on these
observations is rather wide, the cumulative distribution of velocity differences given in Figure 6 
strongly rule out $x_b \gae$ 0.3.
Continued monitoring of the radial velocities of an expanded sample of M4 main-sequence stars will 
clearly provide improved constraints on $x_b$.

These results are summarized as model ``a" in Table 2, whose columns record the 
model ID, the logarithm of the lower and upper period limits (in years), 
the type of period distribution, the minimum and maximum
secondary masses (in solar units), the type of mass ratio distribution 
and the best-fit binary fractions for the case of circular, thermal and halo orbits.
Since the uncertainties given above refer to only {\sl internal} values, we have experimented with
a number of different assumptions regarding the distribution of masses and periods,
the results of which are summarized in the remaining rows of Table 2.
Models b, c and d show the effect of adopting different period and mass ratio distributions,
while the consequences of adopting different minimum secondary masses are explored in models e, f , g and h.
Models i, j and k demonstrate how assumptions concerning the longest possible periods which are detectable 
with the current data affect the derived binary fractions. Finally, model l shows the
effect of making the (unphysical) assumption that all of the binary orbits lie perpendicular to the
plane of the sky.
Clearly, changes caused by varying the model assumptions are 
small compared to the statistical uncertainties in $x_b$.

Our best estimate of $x_b$ $\simeq$ 0.15 for M4 is consistent with the binary fraction for 
nearby, solar-type stars studied by Duquennoy \& Mayor (1991). 
Following the precepts given in \S 2.1.3 of Hut et al. (1992), we find $x_b \simeq$ 0.09 for
binaries in the Duquennoy \& Mayor (1991) survey which have similar periods and mass ratios.
How does our binary fraction compare to other estimates of $x_b$ for M4? The formation
of the hierarchical triple system containing the pulsar PSR 1620-26 has been investigated 
recently by Sigurdsson \& Phinney (1995) who followed the time evolution of a number of 
test binaries in a fixed cluster background. 
These authors concluded that only one such triple --- thought to have
formed through an exchange interaction between a pre-existing binary millisecond pulsar and either
a wide primordial binary composed of two main-sequence stars (Rasio, McMillan \& Hut 1995) or
a main-sequence star and a sub-Jovian mass planet (Sigurdsson 1993) --- is expected per six or seven clusters
with physical parameters similar to M4 {\sl and} having binary fractions of order $x_b \simeq$ 0.5.
In other words, the very existence of the PSR 1620-26 system argues for a rather large binary fraction.
Additional evidence for a non-negligible binary fraction comes from the photometric survey of
Kaluzny (1996), who recently discovered 13 short-period eclipsing binaries in the direction of M4.
Although several of these objects are likely to be field stars since the cluster suffers from considerable
foreground contamination, it is likely that at least {\sl some} of these systems are bona fide cluster 
members --- further evidence that M4 may contain an appreciable number of binaries.

On the other hand, the deep {\sl HST} CMD of Richer et al. (1996) 
shows only marginal evidence for a second sequence. Based on these data, 
Richer et al. (1996) quote a lower limit to the binary fraction of 
$x_b \gae 0.04$.\altaffilmark{7}\altaffiltext{7}{Richer et al. 
(1996) also use the number of probable white-dwarf/red-dwarf pairs in their CMD to derive 
$x_b \sim 0.03$, although this result depends significantly on the assumed background contamination.}
While this estimate applies to binaries of {\sl all} periods, it also refers to 
only those only systems with comparable-mass components. Consequently, it should be compared to the binary 
fraction of $x_b$ $\simeq$ 0.10 derived from our simulations with q = 1 (model h).  
Assuming that binaries with periods longer than $\sim$ 250 years have been disrupted by encounters with other
stars (Hills 1984; C\^ot\'e et al. 1996), this estimate for $x_b$ corresponds to a
sampling of $\sim$ 60\% of the allowed range of orbital periods in M4 (assuming a period distribution which is
flat in $\log{\rm P}$; see below). Thus, the estimate of $x_b$ to be 
compared to that derived from the {\sl HST} CMD is 1.7$x_b$ $\sim$ 0.17.
Given the rather large uncertainties in both estimates, we conclude that the present
results are consistent with those of Richer et al. (1996), although it is worth bearing in mind that the
{\sl HST} results apply to the cluster core ($\simeq$ 1$r_c$), whereas our estimate is based on 
stars near the half-light radius ($\simeq$ 5$r_c$). Sigurdsson \& Phinney (1995) have recently suggested that 
three-body encounters in the dense cores of clusters can lead to a significant overpopulation of binaries
near $\sim$ $r_h$. The existing data for M4 are generally consistent with this picture, although the 
large uncertainties in the derived binary fractions suggest that such conclusions are probably premature.

We now consider how our derived binary fraction compares with those measured for other clusters.
C\^ot\'e et al. (1996) discuss the results of a search for long-period binaries in
the cluster M22. Based on a large number of radial velocities which span more than two decades, 
the best-fit binary fraction for M22 is found to be $x_b$ = 0.01--0.03. In their
survey of $\omega$ Cen, Mayor et al. (1996) find a similarly low binary fraction (i.e., $x_b$ = 0.03--0.04).
Since both of these objects have relatively high binary ionization rates, it is possible that soft
binaries have been disrupted by stellar encounters.
Indeed, surveys which have targeted loose, low-mass clusters, such as 
the one presented here, the survey of NGC 3201 by C\^ot\'e et al. (1994) and the survey of 
M71 by Barden, Armandroff \& Pryor (1996), find considerably higher binary fractions. 
We defer to a companion paper (C\^ot\'e et al. 1996) a more complete discussion of the 
significance and implications of these results
and simply note here that existing radial velocity observations seem to favor 
a binary frequency-period distribution which, for periods in exess of the ``hard-soft" transition (Heggie 1975; Hills 1984),
has been modified by dynamical processes.

\acknowledgments

We thank Tad Pryor for donating the CCD images used to select our program objects. 
PF acknowledges both the Natural Sciences and Engineering Research
Council of Canada (NSERC) and Bell Labs for postdoctoral fellowships. 
Partial support for this work was provided by NASA through grant \# HF-01069.01-94A
from the Space Telescope Science Institute, which is operated by the
Association of Universities for Research in Astronomy Inc., under NASA contract
NAS5-26555.
Additonal support was provided by an NSERC operating grant to Douglas L. Welch, to whom
we extend our thanks.

\vfill\eject

\centerline{~}
\includegraphics{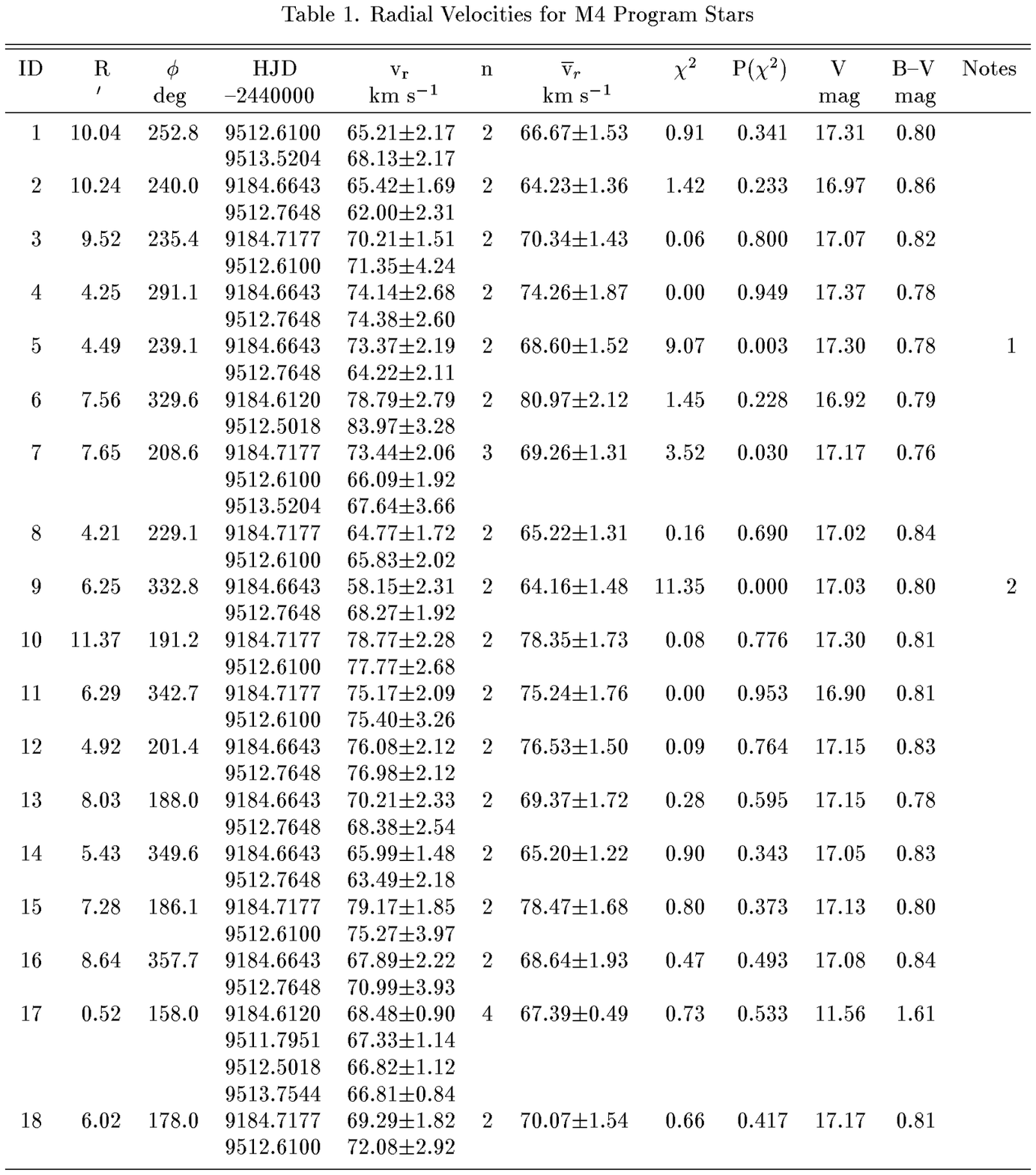}
\vfill\eject
 
\centerline{~}
\includegraphics{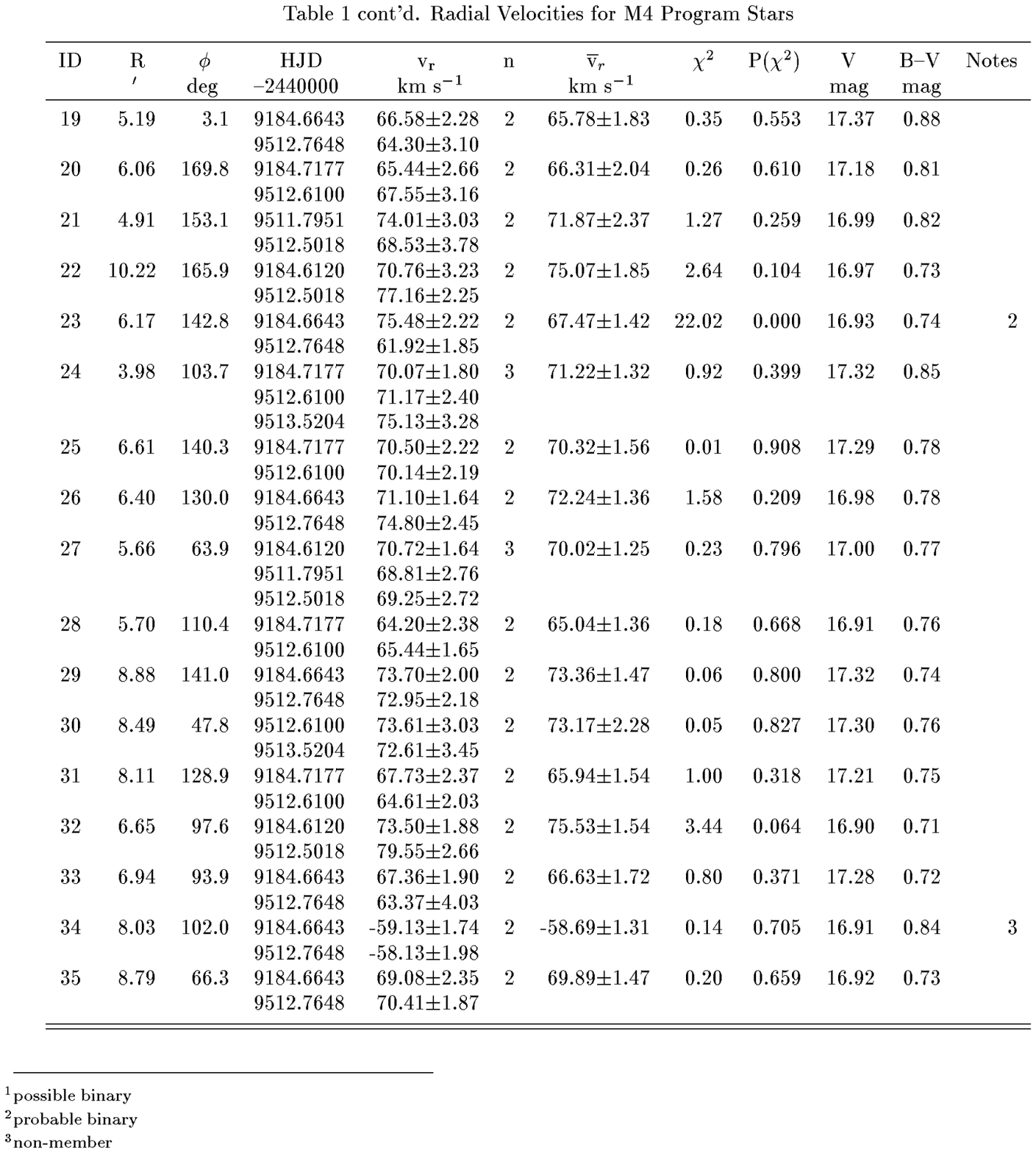}
\vfill\eject
 
\centerline{~}
\includegraphics{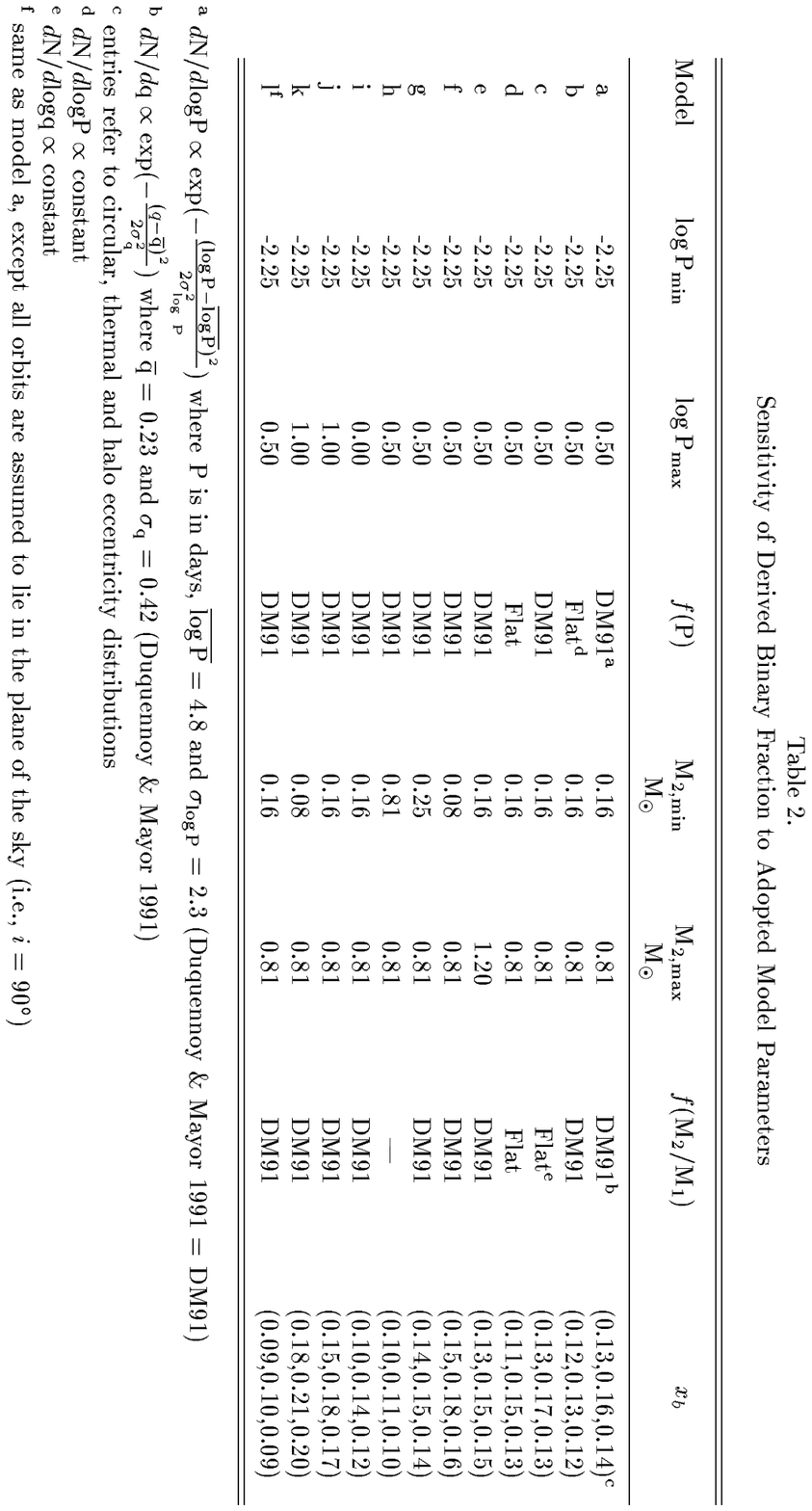}
\vfill\eject

\clearpage

\begin{figure}
\plotone{m4.fig01.ps}
\caption{
{\fsmall
(a) V-band finder chart (11.8$'\times11.8'$) for the NW quadrant of our M4 field. 
Program objects are numbered according to the scheme given
in Table 1. The circular arc indicates a radius of $3.2'$ ($\approx$ 0.7$r_h$ 
according to Trager, King \& Djorgovski 1995).
(b) Same as above, except for SW quadrant.
(c) Same as above, except for SE quadrant.
(d) Same as above, except for NE quadrant.
}
}
\end{figure}

\begin{figure}
\plotone{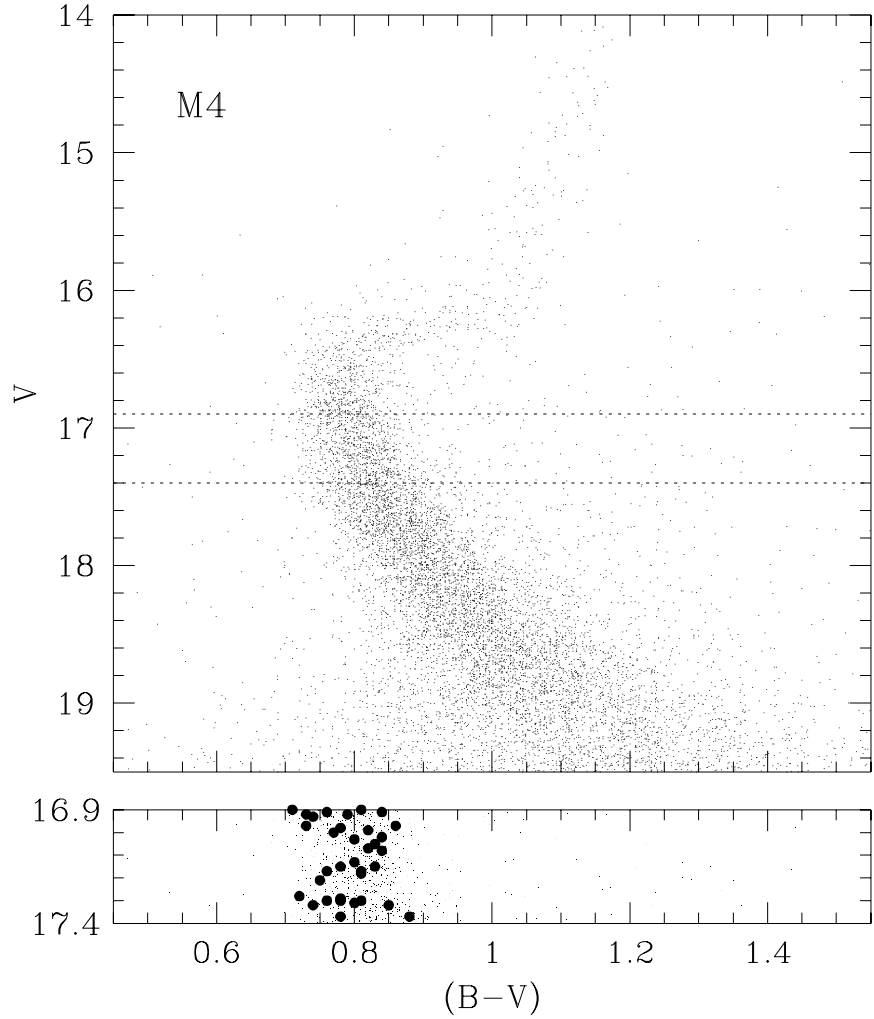}
\caption{
{\fsmall
(Upper Panel) Color-magnitude diagram for M4 based on a pair of BV images taken with the KPNO 0.9m
telescope and 2048$\times$2048 T2KA CCD. The dotted lines indicate the 
magnitude limits (16.9 $\le$ V $\le$ 17.4) of our radial velocity survey.
(Lower Panel) Location in the cluster CMD (i.e., the region enclosed by the dotted lines shown above)
of the 35 program stars having repeat velocities (closed circles).
}
}
\end{figure}

\begin{figure}
\plotone{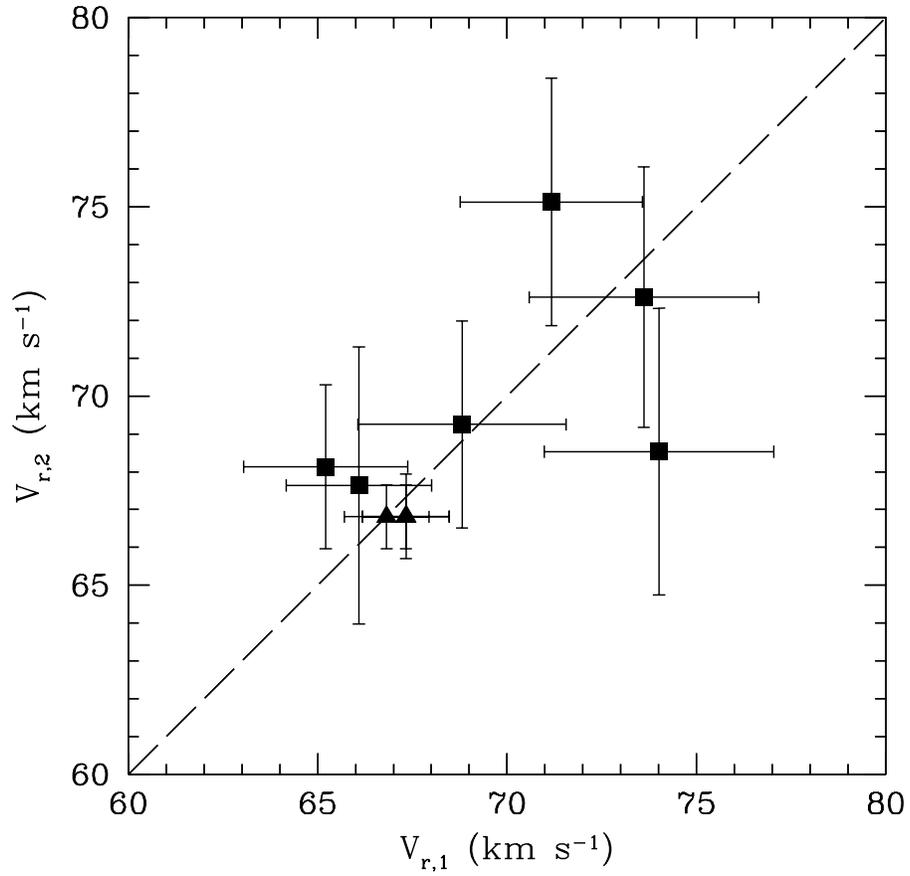}
\caption{
{\fsmall
Comparison of multiple radial velocities for six cluster dwarfs (squares) obtained on different nights
of the June 1994 observing run. The dashed line indicates the one-to-one relation.
Triangles indicate measurements for the lone giant in our survey (star \#4704 of Lee 1977).
}
}
\end{figure}

\begin{figure}
\plotone{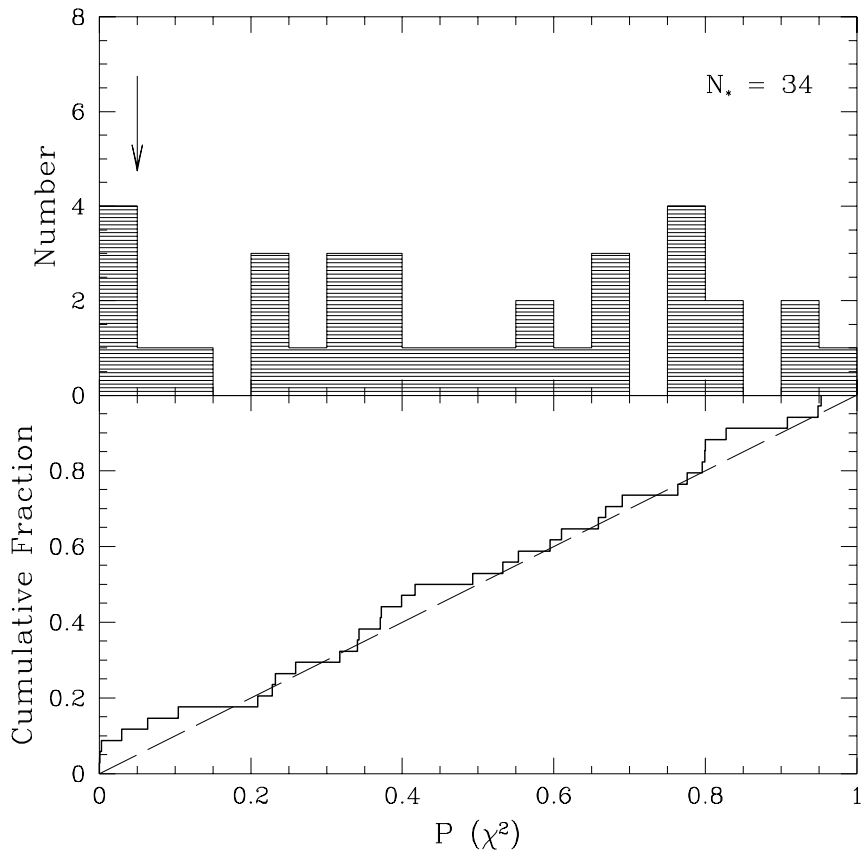}
\caption{
{\fsmall
(Upper Panel) Histogram of P($\chi^2$) for the 34 cluster members having
radial velocities separated by $\simeq$ 11 months. A sample of constant
velocity stars is expected to show a flat distribution, whereas a sample dominated by
radial velocity variables should be strongly peaked at P($\chi^2$) $\approx$ 0. 
The observed distribution is flat, suggesting that we have correctly
estimated our uncertainties. Two stars (\#9 and \#23) have P($\chi^2$) $\le$ 0.001, 
whereas no such objects are expected among a sample of constant velocity stars.
(Lower Panel) Cumulative distribution of P($\chi^2$) for the same 34 stars (solid line).
The dashed diagonal line shows the distribution expected for a sample of constant velocity
stars.
}
}
\end{figure}

\begin{figure}
\plotone{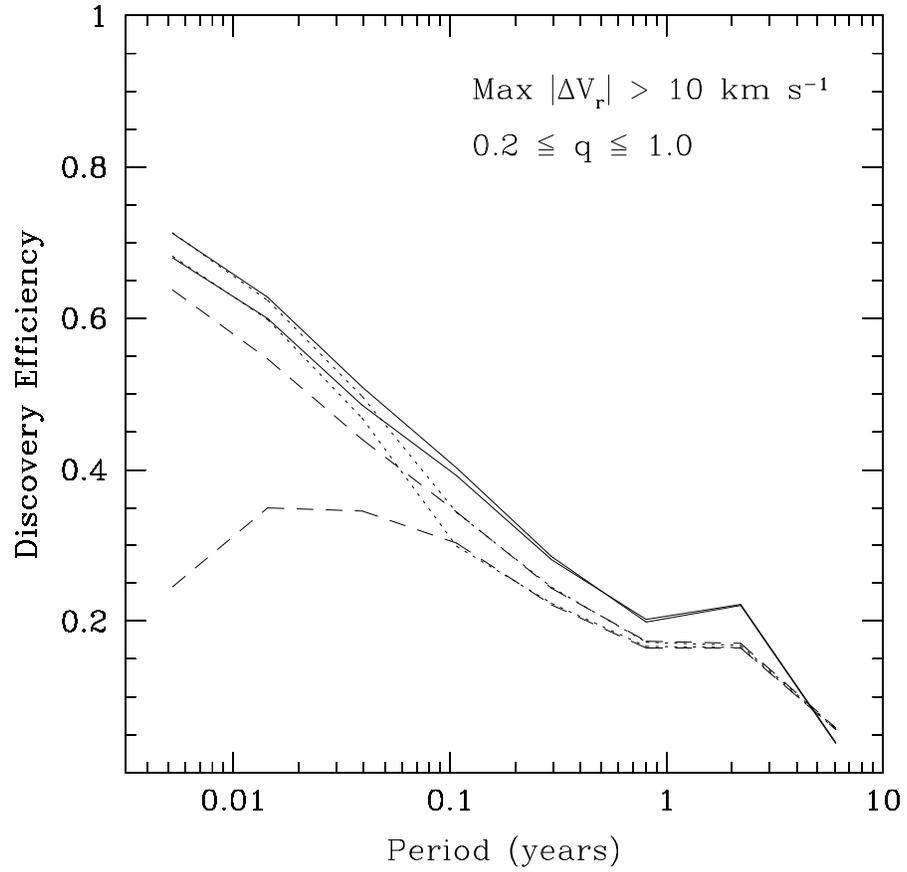}
\caption{
{\fsmall
Binary discovery efficiencies based on the data given in Table 1 for the three cases of
circular orbits (solid lines), a thermal distribution of eccentricities (dashed lines) and
circular orbits for periods below 19 days and a thermal distribution above (dotted lines).
Mass ratios have been randomly chosen from the interval $0.2 \le$ M$_2$/M$_1$ $\le 1.0$.
The upper and lower curves represent the discovery efficiencies before and after taking into
account selection effects caused by Roche-lobe overflow.
}
}
\end{figure}

\begin{figure}
\plotone{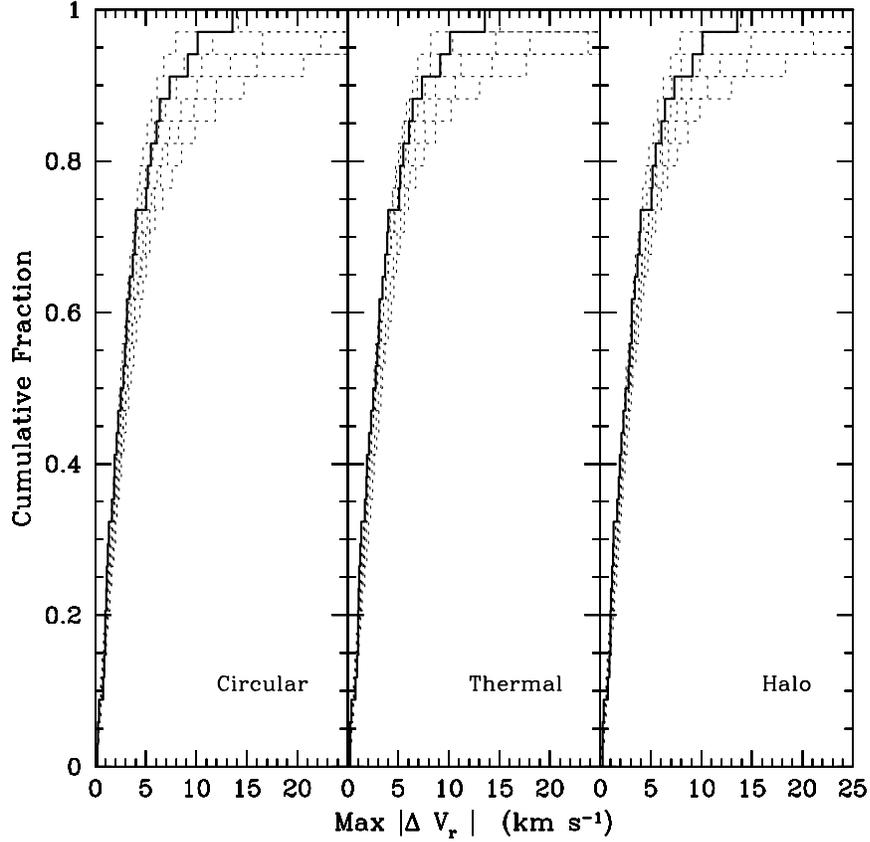}
\caption{
{\fsmall
(Left Panel) Cumulative distribution of maximum radial velocity differences for the
actual data (solid line) and the mean of 100 simulated data sets containing, from top 
to bottom, 0, 8, 16, 24, 32, and 40 percent binaries (dotted curves). The models used 
in producing the simulated distribution functions assume perfectly circular orbits.
(Middle Panel) Same as above, except for a thermal distribution of eccentricities.
(Right Panel) Same as above, except for distribution of eccentricities like that seen in
homogeneous samples of halo binaries. 
}
}
\end{figure}

\end{document}